\documentclass[aps,prl,reprint, amsmath, amssymb,superscriptaddress,nofootinbib]{revtex4-1}

\usepackage{bm}
\usepackage[retainorgcmds]{IEEEtrantools}
\usepackage{graphicx}
\usepackage{mathrsfs}
\usepackage{amsmath}
\usepackage{amssymb}
\usepackage{color}
\usepackage{amsfonts}
\usepackage{bbm}
\usepackage{times,txfonts}
\usepackage{nicefrac}
\usepackage[colorlinks]{hyperref}

\usepackage[dvipsnames]{xcolor}
\definecolor{mypink}{rgb}{0.858, 0.188, 0.478}
\usepackage[authormarkup=none]{changes}
\definechangesauthor[name={Stella}, color=mypink]{s}
\definechangesauthor[name={Stefan}, color=purple]{n}
\definechangesauthor[name={Marti}, color=red]{m}
\definechangesauthor[name={Nicolas}, color=green]{b}
\definechangesauthor[name={Geraldine}, color=blue]{g}

\newcommand{\la}{\langle}
\newcommand{\ra}{\rangle}

\newcommand{\ket}[1]{\ensuremath{\left|#1\right\rangle}}

\newcommand{\ketbra}[2]{\ensuremath{\left| #1 \right\rangle \left\langle #2 \right|}}

\newcommand{\cL}{\mathcal{L}}
\newcommand{\cD}{\mathcal{D}}
\newcommand{\cE}{\mathcal{E}}

\newcommand{\cS}{\mathcal{S}}

\newcommand{\dg}{\dagger}
\newcommand{\da}{\dagger}

\newcommand{\Op}[1]{\hat{#1}}

\newcommand{\osigma}{\Op{\sigma}}

\newcommand{\oH}{\Op{H}}
\newcommand{\oA}{\Op{A}}
\newcommand{\oB}{\Op{B}}

\newcommand{\oU}{\Op{U}}

\newcommand{\oV}{\Op{V}}

\newcommand{\id}{\ensuremath{\mathbbm{1}}}
\newcommand{\tr}{\ensuremath{{\rm tr}}}

\newcommand{\diff}{\mathrm{d}}

\newcommand{\cF}{\mathcal{F}}

\begin{document}

\title{Quantum Speed-up in Collisional Battery Charging}

\author{Stella Seah}
\email{stella.seah@unige.ch} 
\affiliation{D\'{e}partement de Physique Appliqu\'{e}e,  Universit\'{e} de Gen\`{e}ve,  1211 Gen\`{e}ve,  Switzerland}
\author{Mart\'{i} Perarnau-Llobet}
\affiliation{D\'{e}partement de Physique Appliqu\'{e}e,  Universit\'{e} de Gen\`{e}ve,  1211 Gen\`{e}ve,  Switzerland}
\author{G\'{e}raldine Haack}
\affiliation{D\'{e}partement de Physique Appliqu\'{e}e,  Universit\'{e} de Gen\`{e}ve,  1211 Gen\`{e}ve,  Switzerland}
\author{Nicolas Brunner}
\affiliation{D\'{e}partement de Physique Appliqu\'{e}e,  Universit\'{e} de Gen\`{e}ve,  1211 Gen\`{e}ve,  Switzerland}
\author{Stefan Nimmrichter}
\email{stefan.nimmrichter@uni-siegen.de} 
\affiliation{Naturwissenschaftlich-Technische Fakult{\"a}t, Universit{\"a}t Siegen, Siegen 57068, Germany}
\date{\today}

\begin{abstract}
We present a collision model for the charging of a quantum battery by identical nonequilibrium qubit units. When the units are prepared in a mixture of energy eigenstates, the energy gain in the battery can be described by a classical random walk, where both average energy and variance grow linearly with time. Conversely, when the qubits contain quantum coherence, interference effects buildup in the battery and lead to a faster spreading of the energy distribution, reminiscent of a quantum random walk. This can be exploited for faster and more efficient charging of a battery initialized in the ground state. Specifically, we show that coherent protocols can yield higher charging power than any possible incoherent strategy, demonstrating a quantum speed-up at the level of a single battery. Finally, we characterize the amount of extractable work from the battery through the notion of ergotropy. 
\end{abstract}
\maketitle{}

The development of thermodynamic protocols exploiting quantum effects like coherences and entanglement to outperform their classical counterparts has been a major focus in the field of quantum thermodynamics~\cite{Goold2016,Vinjanampathy2016}. A simple scenario to explore this question is a \emph{quantum battery}: a quantum system that receives or supplies energy~\cite{Alicki2013,Campaioli2018,bhattacharjee2020quantum}. Examples include simple qubit batteries~\cite{Horodecki2013,Binder2015}, collective spins~\cite{Andolina2019II}, interacting spin chains~\cite{Le2018,JuliaFarre2020}, and mechanical flywheels~\cite{Levy2016,Roulet2017,Seah2018}. 

Alicki and Fannes developed the first quantum advantage in these devices: entangling operations over multiple batteries can extract more work than local operations~\cite{Alicki2013}. Afterward, the relevance of entanglement for battery charging power was characterized in \cite{Hovhannisyan2013,Binder2015,Campaioli2017}, while implementations were proposed based on collective superradiant coupling in cavity and waveguide QED setups~\cite{Ferraro2018,Ferraro2019,Andolina2019,Pirmoradian2019,Monsel2020}. Speed-ups due to many-body interactions were explored in~\cite{Le2018,Rossini2019,Rossini2020,Rosa2020,Ghosh2020} and theoretical bounds derived in~\cite{JuliaFarre2020}. Charging processes that exploit (or suffer from) dissipation~\cite{Barra2019,Farina2019,Kamin2020,Hovhannisyan2020,Tabesh2020},  stabilization mechanisms~\cite{Santos2019,Gherardini2020,Quach2020,Santos2020,mitchison2020charging}, and the impact of energy fluctuations
\cite{Friis2018,McKay2018,PerarnauLlobet2019,Pintos2020,Crescente2020,Caravelli2020random} have also been investigated.

\begin{figure}
\centering
\includegraphics[width=1.0\columnwidth]{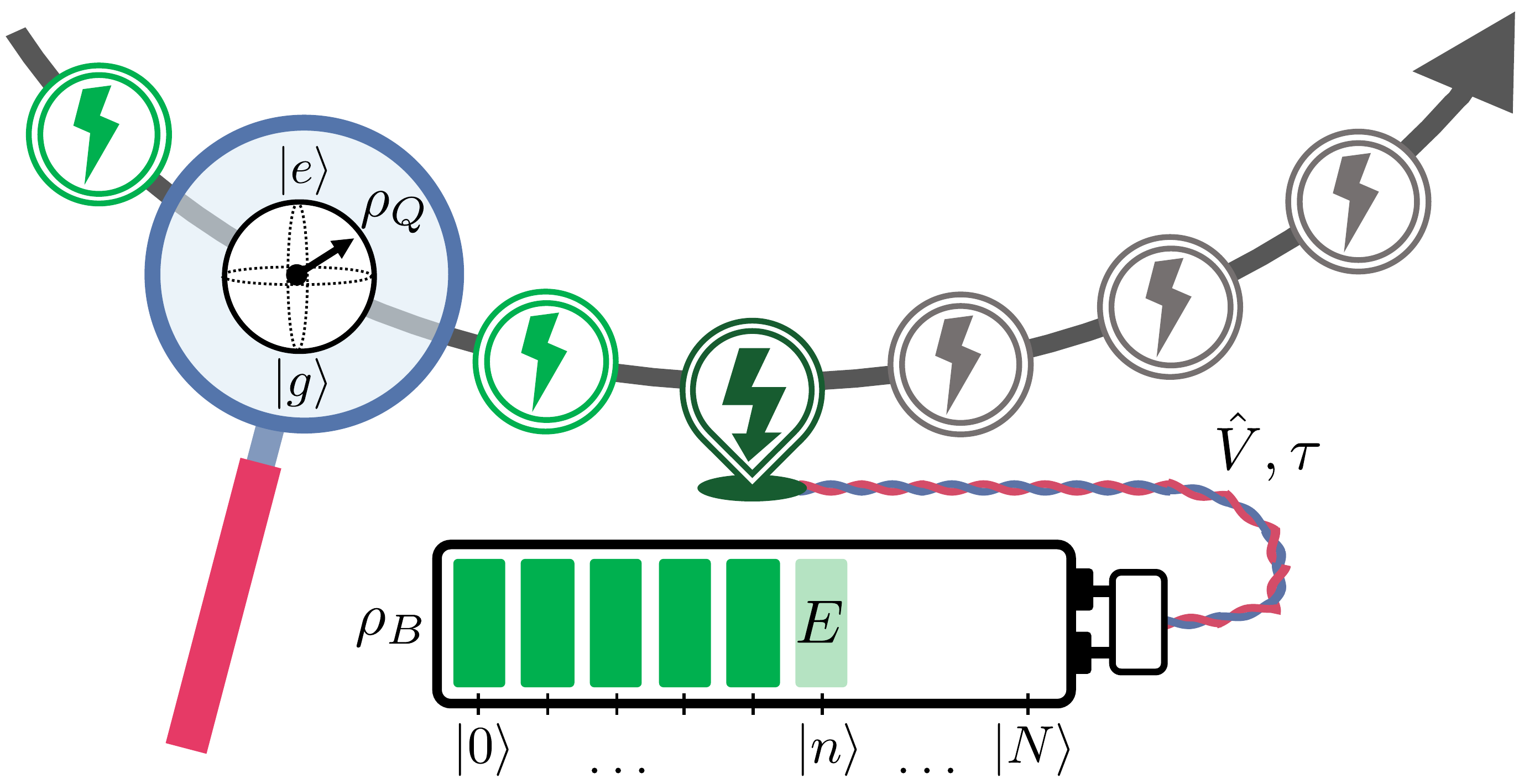}
\caption{\label{fig:sketch} Sketch of the collisional battery charging protocol. The battery is modeled as a uniform energy ladder with $N+1$ levels in steps of $E$. Its quantum state $\rho_B$ receives energy ``charge'' from a stream of identical qubits prepared in the state $\rho_Q$, via the resonant exchange interaction $\oV$ applied for a time $\tau$.
}
\end{figure}

Here we explore the quantum advantage in charging a \emph{single} battery, which arises from the quantum nature of the charging protocol. Any charging process involves auxiliary systems that provide the charge: a thermal engine~\cite{Levy2016,Brunner2012virtual,Bumer2019}, an external time-dependent field~\cite{Campaioli2018,Binder2015,Campaioli2017}, a quantized light field~\cite{Ferraro2018,Ferraro2019,Andolina2019,Pirmoradian2019,Monsel2020}, or more generally an out-of-equilibrium system~\cite{Andolina2019II}. We describe the charging process using a collision model~\cite{Scarani2002,Bruneau2014,Grimmer2016,Strasberg2017,Bumer2019}, see Fig.~\ref{fig:sketch}. Collision models have proven useful to the understanding of equilibration and nonequilibrium dynamics~\cite{Scarani2002,Bruneau2014,Grimmer2016,Seah2019,Cattaneo2021}, the impact of quantum coherence in thermodynamics~\cite{Rodrigues2019,Hammam2021}, as well as strong coupling thermodynamics~\cite{Strasberg2019}.

We find that energy coherences in the charging units generate interference effects in the battery. This leads to a fast spread of the battery's energy distribution, whose variance increases quadratically in time (instead of linear increase with classical units). The behavior is reminiscent of quantum random walks~\cite{Ambainis2001,Kempe2003,Bach2004}, which play a crucial role in quantum algorithms speed-up~\cite{Childs2003,Kendon2006,Childs2007}.
Specifically, we find that coherently-prepared units can lead to faster charging than their classical counterparts for batteries prepared in the ground-state. Notably, when charging power becomes the figure of merit, quantum protocols can overcome arbitrary classical ones.
%at fixed charging time. 
Finally, we study the efficiency of these charging processes, by comparing the ergotropy~\cite{Allah2004} in the battery with that of the charging units. 

\emph{Model.---}We consider a quantum battery with $N+1$ levels, Hamiltonian $\oH_B = E \sum_0^N n \ketbra{n}{n}$, and lowering operator $\oA= \sum_1^N \ketbra{n-1}{n}$ \cite{Brunner2012virtual,Erker2017}. The battery is charged by interacting with a sequence of \emph{resonant} qubits with ground and excited  states $\ket{g},\ket{e}$, and bare Hamiltonian $\oH_Q = E \ketbra{e}{e}$. 

We consider the swap interaction $\oV = \hbar g (\osigma_+\oA +\osigma_-\oA^\dg)$ acting for a time $\tau$ at each charging step, which generates a thermal operation, $[\oV,\oH_B+\oH_Q]=0$ \cite{Brandao2013,Brandao2015}, characterized by a single parameter $\theta = g\tau$. The battery-qubit coupling thus does not require external work~\cite{Barra2015}. 
An arbitrary qubit state,
\begin{equation}\label{eq:qubitState}
  \rho_Q = q \ketbra{g}{g} + (1-q)\ketbra{e}{e} + c \sqrt{q(1-q)} \left( e^{i\alpha}\ketbra{e}{g} + h.c. \right),
\end{equation}
with $q \in [0,1]$ the ground-state occupation, $\alpha \in [0,2\pi]$ the phase, and $c\in [0,1]$ the degree of coherence contains an average energy $E(1-q)$, which is partly transferred to the battery via $\oV$. 

The transformation of the battery state $\rho_B \rightarrow \rho_B'$ after one charge step can be expressed via a Lindblad generator, $\Delta \rho_B = \rho_B' - \rho_B = \cL \rho_B$. 
For a given $q$, we distinguish the two opposing cases of incoherent charging via diagonal qubit states ($c=0$) and coherent charging via pure superposition states ($c=1$). In the rotating frame with respect to $\oH_B + \oH_Q$, we get for $c=0$ the generator
\begin{eqnarray} \label{eq:Lincoh}
 \cL_{\text{inc}} \rho_B &=& \sin^2 \theta \left( q\cD[\oA] \rho_B + (1-q)\cD[\oA^\da]\rho_B \right) \\
 &&+ (1-\cos \theta)^2 \left( q\cD[ |0\ra\la 0|] \rho_B + (1-q)\cD[|N\ra\la N|]\rho_B \right),\nonumber 
\end{eqnarray}
where $\cD[\oB]\rho = \oB\rho\oB^\da - \{\oB^\da \oB,\rho\}/2$.
The first line describes jumps up and down the energy ladder, conditioned on the qubit's ground ($q$) and excitation probabilities ($1-q$). This constitutes a random walk with an overall jump probability $p_\theta = \sin^2 \theta$ and leads to an average energy change per charge step of
\begin{equation}\label{eq:battEnergy}
  \tr \{ \oH_B \cL_{\rm inc} \rho_B \} = v E + p_\theta E \left[ q \la 0|\rho_B|0\ra - (1-q) \la N|\rho_B |N\ra \right].
\end{equation}
Here, $v=p_\theta (1-2q)$ denotes the (classical) dimensionless drift across the battery. % per charge step. 
Average energy growth ($v>0$) requires population inversion, $q<1/2$. For a finite battery, the jumps are terminated at the boundaries and complemented by dephasing of the boundary states, as described in the second line of \eqref{eq:Lincoh}: 
if no jump occurs, a ground-state (excited) qubit would learn that the battery is empty (full). The dephasing would destroy any initial energy coherence.

For $c=1$, we get
\begin{eqnarray} \label{eq:Lcoh}
 \cL_{\text{coh}} \rho_B &=& -i \sqrt{q(1-q)}\sin\theta \cos\theta \left[ \oA e^{-i\alpha} + \oA^\da e^{i\alpha} , \rho_B \right] \\
 &&+ \cD \left[ \sqrt{q}\sin\theta \oA + ie^{i\alpha}\sqrt{1-q}(1-\cos\theta) |N\ra\la N| \right] \rho_B \nonumber \\
 &&+ \cD \left[ \sqrt{1-q}\sin\theta \oA^\da + ie^{-i\alpha}\sqrt{q}(1-\cos\theta) |0\ra\la 0| \right] \rho_B.\nonumber
\end{eqnarray}
The coherent generator \eqref{eq:Lcoh} contains all incoherent terms with additional cross-terms and an effective coherent driving Hamiltonian (first line). This driving term can generate interference effects in the case of strictly \emph{partial} swaps ($\sin 2\theta \neq 0$). We identify the effective Rabi parameter
\begin{equation} \label{eq:Rabi}
  \Omega := 2\sqrt{q(1-q)}\sin\theta \cos\theta = \sqrt{q(1-q)} \sin 2\theta,
\end{equation}
which quantifies the coherent speedup of battery charging. Other $c-$values would result in a mixture of the two generators, $\cL \rho_B = c \cL_{\text{coh}} \rho_B + (1- c ) \cL_{\text{inc}} \rho_B$.

\emph{Incoherent charging.---}We first consider a classical charging protocol ($c=0$). From \eqref{eq:Lincoh}, the battery state then remains diagonal, fully characterized by the populations $P(n,k) = \la n|\rho_B(k) | n\ra$ after interaction with the $k-$th qubit. In the regime where the battery does not populate the boundary states $\ket{0}$ and $\ket{N}$, 
we obtain a discrete three-branch random walk with an update rule
\begin{equation}
  P(n,k+1) = (1-p_\theta) P(n,k) + p_\theta (1-q) P(n-1,k) + p_\theta q P(n+1,k), \label{eq:classRW}
\end{equation}
valid for $0<n<N$. 
The variable for the number of jumps up/down or no jumps after $k$ interactions, $\mathbf{X} = \left\{X_{+},X_{-},X_0\right\}$, follows a trinomial distribution with probabilities $\mathbf{p} = \left\{p_\theta (1-q), p_\theta q, 1-p_\theta\right\}$ \cite{Franke2015}. The mean and variance of charge~$n$ (energy in units $E$) grow linearly in time, 
\begin{equation}\label{eq:incohMeanEnergy}
 \overline{n}(k) = \overline{n}(0) + v k, \qquad \Delta n^2 (k) = \Delta n^2 (0) + (p_\theta - v^2)k,
\end{equation} 
with $\overline{n}(k)=\sum_n n P(n,k)$ and $\Delta n^2 (k)=\sum_n n^2 P(n,k) - \overline{n}(k)^2$.
Taking $k\gg 1$, the trinomial distribution has two natural limits.
For small jump probabilities ($p_\theta \ll 1$) and finite $k p_\theta$, the number of jumps up/down obey a Poisson distribution with mean $k p_\pm$~\cite{Lawler2010,Deheuvels1988}. Conversely, for large $k$ and moderate $p_\theta\not\approx0,1$, $\mathbf{X}$ follows a Gaussian distribution in accordance with the central limit theorem \cite{Supp}. The battery populations eventually converge to $P(n,\infty)\propto(1-q)^n/q^n$, the fixed point of \eqref{eq:classRW}. It amounts to a Gibbs state at an effective (negative) qubit temperature defined by $e^{-E/k_B T} = (1-q)/q$; this follows by noting that the product Gibbs state $\rho_Q \otimes\rho_B \propto e^{-(\oH_Q + \oH_B)/k_B T}$ is invariant under $\oV$ while population inversion ($q< 1/2$) implies~$T<0$.

\emph{Coherent charging and quantum signatures.---}Now, instead of using classical population-inverted states to charge the battery, consider qubits prepared in pure superposition states with the same occupation $q$. They have the same mean energy but
a higher purity, which can be interpreted as a thermodynamic resource \cite{Rodrigues2019,Hammam2021}. Since $\oV$ is invariant under rotation about the $z-$axis, we set the qubit phase $\alpha = 0$ without loss of generality. 

Ignoring boundary effects, one can show that a bimodal energy distribution emerges. The two branches simultaneously progress down and up the energy ladder with increasing $k$~\cite{Supp}. 
For an initially pure battery state $\ket{n_0}$ and sufficiently large $k$ ($n_0-k>0$ and $n_0+k<N$), the two branches has energy peaks at approximately
\begin{equation}\label{eq:cohPeaks}
  n_{\pm} \approx n_0 + ( v \pm c \Omega ) k
\end{equation}
%on the energy ladder. 
The coherent driving thus speeds up (and slows down) each peak by $c\Omega$, resulting in the same mean energy increase as in the incoherent case \eqref{eq:incohMeanEnergy}. 
However, the energy variance now grows quadratically with $k$, 
$\Delta n^2 (k) \approx (p_\theta - v^2) k + c^2 \Omega^2 k(k-1)/2$, which can be seen as a genuine quantum signature.

\begin{figure}
\centering
\includegraphics[width=\columnwidth]{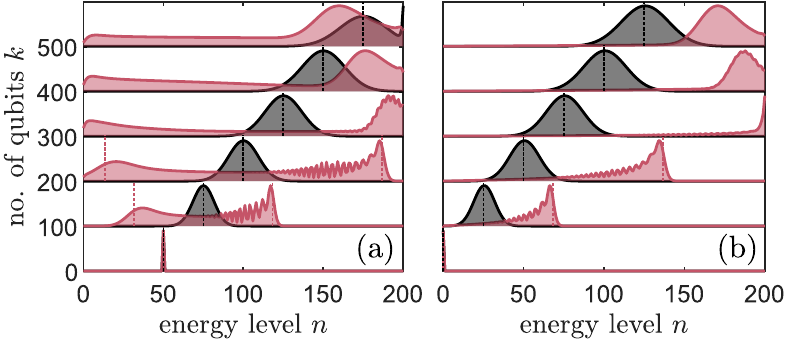}
\caption{\label{fig:distribution} Energy distributions of a battery of size $N=200$ charged by $k$ incoherent (black) and coherent (red) qubits with $q=1/4$ and $\theta = \pi/4$. All curves are rescaled to the same maximum. In (a), the battery is initialized in the pure, partially charged state $\ket{n_0=50}$, whereas in (b), the battery is initialized in the zero-charge state $\ket{0}$. The dotted black and red vertical lines mark the approximate average charge according to \eqref{eq:incohMeanEnergy} and the approximate peak positions \eqref{eq:cohPeaks} for incoherent and coherent qubits, respectively for $\overline{n}<200$.}
\end{figure}

Exemplary snapshots of the battery's energy distribution with $k$ are depicted in Fig.~\ref{fig:distribution}(a) for coherent (red) and incoherent charging (black). The incoherent case is well-described by Gaussians whose mean values (black-dotted lines) move according to \eqref{eq:incohMeanEnergy}. The red-dotted lines mark the approximate peak positions \eqref{eq:cohPeaks} for the coherent case, which agree with the actual distribution maxima. Note that the average energy $\overline{n}(k)$ is the same in both cases as long as the boundaries are not hit. Once this happens, the quantum wave is reflected, whereas the classical distribution approaches the inverse Gibbs state. 

\emph{Quantum advantage for empty batteries}.---At first sight, the previous considerations suggest that quantum coherence is detrimental for battery charging: coherent qubits provide at most the same average energy to the battery as incoherent ones, but with a larger variance. 
The situation drastically changes when the battery is initially empty ($n_0=0$). Now the charge distribution comprises only a single forward-propagating peak, see Fig.~\ref{fig:distribution}(b). 
Even though our analytic random-walk model no longer applies, we verify numerically that the peak is still situated close to ~$n_+$ from~\eqref{eq:cohPeaks}, which climbs up the ladder at a 73\% faster rate than the incoherent Gaussian for $q=0.25$. Coherence results in interference effects and directly attributes to the speed-up by $c\Omega$. In fact, both purity and coherence can be seen as independent resources for this task: a higher purity leads to a higher (classical) drift velocity~$v=p_\theta (1-2q)$, whereas coherence adds an additional contribution   $c\Omega$ to the velocity (see  Eq~\eqref{eq:cohPeaks}). 

\begin{figure}
\centering
\includegraphics[width=\columnwidth]{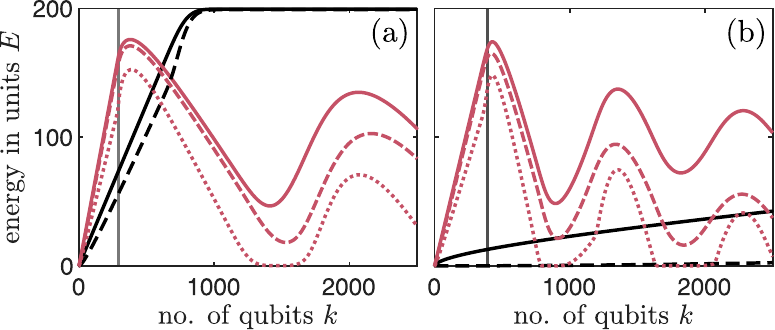}
\caption{\label{fig:energy} Energy (solid) and ergotropy (dashed) for incoherent (black) and coherent (red) charging of a battery of $N=200$ with $k$ qubits. Here, (a) matches Fig.~\ref{fig:distribution}(b) with $q=0.25$ and $n_0 = 0$ while (b) corresponds to $q=0.49$. The dotted line shows the ergotropy associated to the dephased battery state, i.e.~only the energy populations. The vertical lines mark the theoretical $k_{\rm est}$.}
\end{figure}

We compare the energy growth of incoherent and coherent protocols in Fig.~\ref{fig:energy}(a) for $q=0.25$; and in (b) for the more distinctive case $q=0.49$. 
Far below the fully-charged state, 
coherent charging is always more advantageous. However, once the maximum charge is reached, we see a drop in energy due to reflection at the boundary $n=N$. The number of coherent steps from empty to maximum battery charge is therefore 
\begin{equation}\label{eq:kest}
    k_{\rm est } \approx \left\lfloor \frac{N}{c \Omega + v} \right\rfloor 
\end{equation}

For Fig.~\ref{fig:energy}, this amounts to (a) $k_{\rm est} \approx 292$ and (b) $392$, in good agreement with the depicted maxima. Beyond this point, the quantum protocol loses its advantage, whereas the incoherent protocol continues to charge. 
This suggests a hybrid protocol in which the battery is charged coherently until $k\approx k_{\rm est}$ to exploit the quantum speedup, then incoherently 
until the battery reaches an inverted Gibbs state.

Focusing on $k<k_{\rm est}$, the quantum advantage can be quantified by 
the ratio between incoherent ($c=0$) and coherent ($c=1$) energy charging rate:
\begin{align}
  \mathcal{A}(k)=\frac{\overline{n}_{c=1}(k)-\overline{n}_{c=1}(k-1)}{\overline{n}_{c=0}(k)-\overline{n}_{c=0}(k-1)}%\frac{\Delta n_{c=1}(k)}{\Delta n_{c=0}(k)}
\end{align}
Here the subindex in $\overline{n}(k)$ distinguishes both cases.
There is a quantum advantage if $ \mathcal{A}>1$ for some $k$, $q$ and~$\theta$.
In the regime $N>k\gg 1$, we obtain an upper-bound on $\mathcal{A}$ by noting that the increment of $\overline{n}_{c=1}$ is upper-bounded by the peak velocity
$v+\Omega$, whereas the classical increment is, for the most part, given by $v$,
\begin{align}
\label{eq:uppboundqa}
  \mathcal{A} \lesssim 1+ \frac{\Omega}{v} = 1+\frac{2\sqrt{q(1-q)}}{ (1-2q)\tan{\theta}}.
\end{align}
It can be verified numerically that $\mathcal{A}$ is always above 70$\%$ of this bound. Moreover, the quantum advantage diverges for $q\rightarrow 1/2$ and for $\theta\ll 1$. This can be understood by noting that for $q\approx 1/2$, incoherent charging vanishes while coherent charging is still possible. For short interactions ($\theta \ll 1$), 
the coherent driving term in \eqref{eq:Lcoh} is first-order in $\theta$, while the classical charging terms are of second-order. 

\emph{Ergotropy and charging efficiency.---}Due to inevitable energy fluctuations, not all energy stored in the battery is useful work. This effect is well captured by the notion of ergotropy~\cite{Allah2004,Goold2016}, which characterizes the maximum amount of useful energy as the part that can be extracted by means of a deterministic unitary operation. The unitary brings the battery to a passive state~\cite{Pusz1978,Lenard1978,Allah2004}, given by the energy mixture $\pi_B = \sum_{0}^N r_n |n\ra\la n|$ with $\{r_n\}$ the eigenvalues of the initial state in \emph{descending} order.

Fig.~\ref{fig:energy} shows the ergotropy and average energy of the battery. 
When charged by coherent qubits, the battery stores useful energy in two ways: by populating higher energies and building up energy coherences. We distinguish the two by considering the residual ergotropy from a dephased battery state in the energy basis (dotted).
In fact, most useful energy is contained in the energy profile during the charging of an initially empty battery until it reaches maximum charge. In the long-time limit, however, ergotropy will approach comparatively low values almost entirely contained in coherences: for $c=1$ in Fig.~\ref{fig:energy}(a) and (b), the ergotropy converges to $51E$ and $14E$ at $k \to \infty$, respectively, while the dephased-state ergotropy converges to $1E$ and $0.2E$.

Ergotropy provides a natural way to define the charging efficiency without a reference temperature \cite{Supp}. For $k$ qubits, each with ergotropy $\cE_Q= E \left[1-2q + \sqrt{(1-2q)^2 + 4c^2q(1-q)} \right]/2$, we define the efficiency as the ratio of stored ergotropy over total ergotropy input, $\eta(k) = \cE_B(k)/k \cE_Q$.
Note that $\eta(k)\leq 1$ for energy-preserving interactions \cite{PerarnauLlobet2019}. 

\begin{figure}
\centering
\includegraphics[width=\columnwidth]{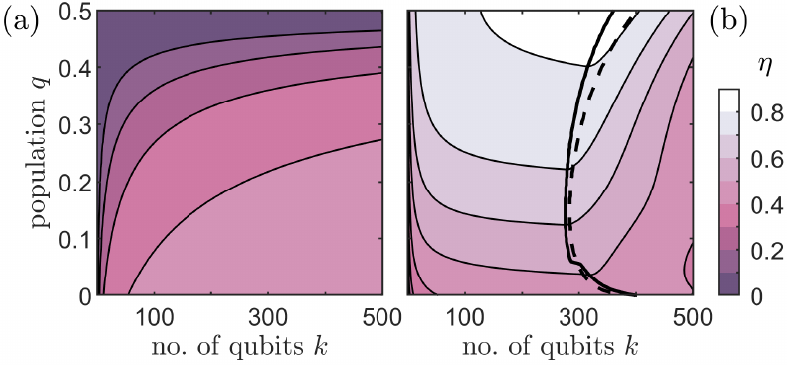}
\caption{\label{fig:compareEff} (a) Incoherent and (b) coherent charging efficiency $\eta$ for a battery of size $N=200$ initialized in the zero-charge state for different $k$ and $q$. Here, we consider $\theta = \pi/4$ where interference effects are the strongest. 
The solid line in (b) marks the number of qubits that maximises $\eta$ for a given $q$-value, while the dashed line marks the estimated $k_{\rm est}$ in \eqref{eq:kest}.}
\end{figure}

Figure \ref{fig:compareEff} shows the (a) incoherent and (b) coherent charging efficiency as a function of $k$ and $q$. Coherent ergotropy transfer can reach more than 80\% maximum efficiency at intermediate qubit numbers $k$, indicated by the solid curve in Fig.~\ref{fig:compareEff}(b) and compared against $k_{\rm est}$ (dashed). The efficiency drops for greater $k-$values after reflection at the full-charge state $|N\ra$. Incoherent charging slows down approaching the inverted Gibbs state.

\emph{Quantum advantage in power.---}The previous considerations show that quantum coherence can be exploited to enhance battery charging and its efficiency %$n(k)$
given a set of $k$ qubits with fixed interaction $\theta$. 
However, the energy $\overline{n}(k)E$ charged into the battery would be maximum for perfect population inversion ($q\rightarrow 0$), and charging efficiency would reach $\eta(k)=1$ for full swaps ($\theta=\pi/2$). Crucially, this best-case-scenario does not require quantum coherence, which raises the natural question: can we find a quantum advantage against arbitrary classical protocols?
Remarkably, the answer is affirmative when charging \emph{power} is the figure of merit and \emph{time} the only constraint.

\begin{figure}
\centering
\includegraphics[width=\columnwidth]{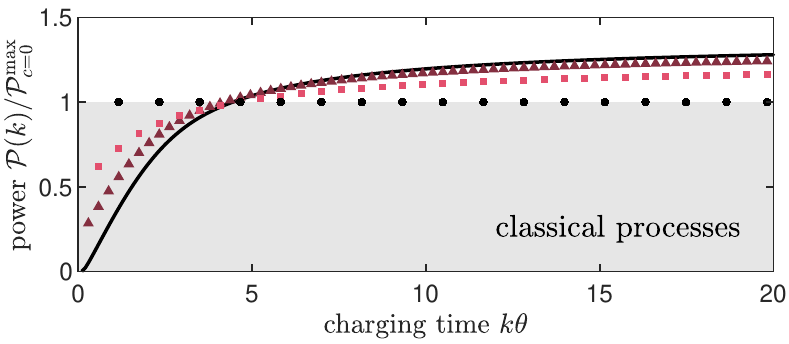}
\caption{\label{fig:quantadpower} Quantum advantage in terms of power normalized to the best incoherent charging protocol (circles), which uses a sequence of excited qubits ($q=0$) with interaction parameter $\theta_{\max}$. The gray region bounds all classical processes. The solid curve represents the continuous quantum limit using $k\to \infty$ coherent qubits at $q=1/2$. The squares and triangles represent coherent protocols for finite qubits with $q=0.26$ and $0.38$ interacting over $\theta = \theta_{\max}/2$ and $\theta_{\max}/4$, respectively. The chosen $q$-values yield optimal power in the regime of $N>k\gg 1 $. }
\end{figure}

Starting with an empty battery ($\overline{n}(0)=0$), the average charging power of $k$ qubits, $\mathcal{P}(k) = g E \overline{n}(k)/k\theta = E \overline{n}(k)/t$, quantifies the gain in average battery energy over the charging time $t = k\theta/g$.
We shall maximise $\mathcal{P}$ over $k$ and $q$ at \emph{fixed} $t$.
For incoherent probes, power decreases monotonically with $q$, and the optimal strategy is hence obtained for pure excited qubit states at $q=0$. A straightforward calculation yields the fundamental upper-bound for incoherent charging, $\mathcal{P}_{c=0}^{\max} \approx 0.62 gE$ attained at $\theta_{\max} \approx 1.17$.

Coherent probes can overcome this bound, see Fig.~\ref{fig:quantadpower}, which singles out coherence (and not purity) as the relevant thermodynamic resource causing the advantage.That is, coherent qubits ($c\neq 0$) can outperform arbitrary incoherent strategies, including pure excited qubits ($q=0$).

The optimal coherent strategy involves taking $q=1/2$ and the limit of $\theta \to 0$ and $k\to\infty$, which caps the velocity of the fast-moving peak in~\eqref{eq:cohPeaks} at $\Omega k/t \to g$ (solid line). In this case, the power is initially smaller, but after a transient buildup of quantum coherence, it overcomes the incoherent bound (circles) and hits $\mathcal{P}_{c=1}^{\max} \approx 0.85 gE = 1.37 \mathcal{P}_{c=0}^{\max} $ at $1\ll k < N$.  In practice, one can already achieve a significant coherent power enhancement with finite qubits (triangles and squares). Importantly, the advantage remains in terms of ergotropy. At $k\theta = 20$, the ergotropy charging power still exceeds the classical bound, and amounts to 91\% and 96\% of $\mathcal{P}_{c=1} (k)$ for $\theta_{\rm max}/2$ and $\theta_{\rm max}/4$, respectively. In the limit $q=1/2$ and $\theta\to 0$, the ergotropy matches exactly the energy gain since the battery evolves almost unitarily, building up energy in its coherences through $\oA + \oA^\dg$ according to \eqref{eq:Lcoh}.

\emph{Conclusions.---}We proposed a collision model for quantum battery charging, in which %a set of
identical qubits (charging units) transfer energy to a finite energy ladder (battery). We compared classical to quantum charging protocols where the units are prepared in a mixture or superposition state, respectively. 
We found that quantum protocols can yield a higher power than arbitrary classical strategies, thus providing a clear advantage at the level of a single battery. This complements previous examples of collective quantum speed-ups~\cite{Campaioli2018,bhattacharjee2020quantum}. Our analysis also highlights a connection between quantum thermodynamics and quantum random walks~\cite{Ambainis2001,Kempe2003,Bach2004}. The formalism behind is however different: in quantum random walks a single coin becomes entangled with the walker, whereas here interference effects in the walker (battery) are created through partial collision processes with numerous coins (units). Future work could further reveal the connection between both frameworks and exploit quantum walk-like features in thermodynamic protocols.

\begin{acknowledgments}
S.S. and N.B. acknowledge funding from Swiss National Science Foundation (NCCR SwissMAP). M. P.-L. and G.H. acknowledge funding from Swiss National Science Foundation through an Ambizione grant PZ00P2-186067 and a starting grant PRIMA PR00P2-179748 respectively. 
\end{acknowledgments}

\clearpage

\appendix
\begin{widetext}

%\section{Infinite ladder charging by arbitrary spin states}\label{app:infLadderCharging}
\section{Charging a quantum battery by a qubit}\label{app:infLadderCharging}

We model the quantum battery as a regular energy ladder of equidistant levels $|n\ra$ separated by the energy gap $E$. It shall be charged in discrete steps of resonant energy exchange with identical qubits prepared in arbitrary quantum states $\rho_Q$, as mediated by the energy-preserving unitary 
\begin{equation}\label{eq:map}
 \oU_\theta = \exp \left[-i\theta \left( \oA |e\ra\la g| + \oA^\da |g\ra\la e|  \right) \right],
\end{equation}
and its inverse $\oU^\da_\theta = \oU_{-\theta}$. Here, $|g\ra,|e\ra$ denote the qubit's ground and excited state and $\oA,\oA^\da$ denote the lowering and raising operators on the battery ladder. The reduced battery state transforms as $\rho_B \to \rho_B' = \tr_Q \{ \oU_\theta \rho_B \otimes \rho_Q \oU_{-\theta} \}$. 

For an idealized, infinite battery, the ladder operators are defined through $\oA|n\ra = |n-1\ra$ and $\oA^\da |n\ra = |n+1\ra$ for all $n$, which leads to
\begin{equation}\label{eq:U_matrix}
    \oU_\theta |n,g \ra = \cos \theta |n,g \ra - i\sin \theta |n-1,e \ra, \quad \oU_\theta |n,e \ra = \cos \theta |n,e \ra - i\sin \theta |n+1,g \ra .
\end{equation}
Given a generic qubit state of the form
\begin{equation}
    \rho_Q = q |g \ra \la g| + (1-q) |e \ra \la e| + c \sqrt{q(1-q)} \left(  e^{-i\alpha} |g \ra \la e| + e^{i\alpha} |e \ra \la g| \right), \label{eq:qubitState}
\end{equation}
a straightforward application of \eqref{eq:U_matrix} shows that the transformation of the battery state after one step is
\begin{align}
    \la n |\rho_B' | n'\ra &= (1-p_\theta) \la n |\rho_B|n'\ra + p_\theta \left[ (1-q) \la n-1 |\rho_B|n'-1\ra + q \la n+1 |\rho_B|n'+1\ra \right] \nonumber \\
    &- i c \frac{ \Omega}{2}  \left[ e^{i\alpha} \left( \la n-1 |\rho_B|n'\ra - \la n |\rho_B|n'+1\ra \right) + e^{-i\alpha} \left( \la n+1 |\rho_B|n'\ra - \la n |\rho_B|n'-1\ra \right) \right]. \label{eq:qubitChargeTrafo}
\end{align}
in the energy representation, with $p_\theta = \sin^2 \theta$ and $\Omega = \sqrt{q(1-q)}\sin 2\theta$. This transformation, which describes the time evolution of an infinite battery in discrete charging steps, can generate a classical or quantum random walk process depending on the chosen parameters $\theta,q,\alpha,c$. In the case of a finite battery with $N+1$ levels from the zero-charge state $|0\ra$ to the fully charged $|N\ra$, Eqs.~\eqref{eq:U_matrix} and \eqref{eq:qubitChargeTrafo} still hold for any $0 < n,n'< N$. Hence the infinite-battery results remain valid for those battery states that do not occupy $|0\ra$ or $|N\ra$. 

In order to obtain an expression for the transformation of finite-battery states, we can expand the unitary \eqref{eq:map} as
\begin{equation}\label{eq:map_finite}
 \oU_\theta = \cos\theta \, \id - i\sin\theta \left( \oA \otimes |e\ra\la g| + \oA^\da \otimes |g\ra\la e| \right) + (1-\cos\theta) \left( |0,g\ra\la 0,g| + |N,e\ra\la N,e| \right) ,
\end{equation}
with $\oA = \sum_{n=1}^N |n-1\ra\la n|$. Here, the last term accounts for the modified effect at the charge boundaries. 
The final state is given by $\rho_B' = \tr_Q[\oU_\theta \left(\rho_B\otimes \rho_Q\right) \oU_\theta ^\dg]$ with $\rho_Q$ defined in \eqref{eq:qubitState}. 
Plugging $\rho_Q$ into $\rho_B'$ and reordering the terms, one can write the incremental change in the reduced battery state $\Delta \rho_B = \rho_B' - \rho_B = \cL \rho_B$ in terms of the Lindblad generator $\cL$ defined in the main text.
%in Eq.~\eqref{eq:L} in the main text.

In the following, we study the time evolution of the battery state $\rho_B(k) = (\id + \cL)\rho_B(k-1) = (\id +\cL)^k \rho_B(0)$ after $k$ charge steps, omitting boundary effects. We distinguish the opposite cases of incoherent charging ($c=0$) and coherent charging ($c=1$).

\section{Incoherent battery charging}\label{app:infLadderClassical}

Assuming an initially diagonal battery state and diagonal qubits ($c=0$), we are left with a classical discrete-time random walk model. The battery state remains diagonal at all times, and the populations transform according to a simple Markov chain with three branches: no jump with probability $1-p_\theta$, jump up with $p_\theta (1-q)$, jump down with $p_\theta q$. Denoting by $P(n,k) = \la n|\rho_B (k) | n\ra$ the populations of the battery at discrete time steps $k=0,1,\ldots$, we get from \eqref{eq:qubitChargeTrafo},
\begin{equation}
    P(n,k+1) = (1-p_\theta) P(n,k) + p_\theta (1-q) P(n-1,k) + p_\theta q P(n+1,k). \label{eq:classRWprop}
\end{equation}
This difference equation is \emph{exact} for an infinite ladder as well as for a finite ladder at $0<n<N$, which implies that its solution may serve as an approximation for transient states of finite ladders so long as the population at the charge boundaries is small. Specifically, the first and second moments and cumulants of charge $n$ evolve as
\begin{align}\label{eq:classRWmoments}
    \overline{n}(k) &= \sum_{n=-\infty}^\infty n P(n,k) = \overline{n}(k-1) + p_\theta (1-2q) = \overline{n}(0) + p_\theta (1-2q) k =: \overline{n}(0) + v k, \nonumber \\
    \overline{n^2}(k) &= \overline{n^2}(k-1) + p_\theta + 2 p_\theta (1-2q) \overline{n}(k-1) = \overline{n^2}(k-1) + p_\theta + 2 v \overline{n}(k-1), \\
    \Delta n^2 (k) &= \overline{n^2}(k) - \overline{n}^2(k) = \Delta n^2 (k-1) + p_\theta - v^2 = \Delta n^2 (0) + (p_\theta - v^2) k, \nonumber 
\end{align}
indicating that the charge distribution spreads and drifts, as described by the stepwise increment $v = p_\theta (1-2q)$ of the mean charge. The time-evolved $P(n,k)$ can be expressed analytically in terms of the initial $P(n,0)$ in various ways. 

For example, one can view the $P(n,k)$ as the $n$-th discrete Fourier coefficients of a periodic characteristic function $\chi (\phi,k) = \sum_n P(n,k) e^{in\phi}$ with $\phi \in (-\pi,\pi]$, normalized to $\chi(0,k) = 1$. The Markov chain \eqref{eq:classRWprop} translates into
\begin{align}
    \chi(\phi,k+1) &= \left[ (1-p_\theta)  + p_\theta (1-q) e^{i\phi} + p_\theta q e^{-i\phi} \right] \chi(\phi,k) \quad 
    \Rightarrow \quad \chi(\phi , k) = \left[ 1 - p_\theta \left( 1 - \cos \phi \right) + i v \sin\phi \right]^k \chi(\phi,0) 
    . \label{eq:chiSol_class}
\end{align}
This amounts to a geometric progression by a factor of magnitude smaller than unity. Indeed, one can easily check that the magnitude of the square-bracketed expression is never greater than $\sqrt{1-4p_\theta(1-p_\theta)\sin^2(\phi/2)}$ for any $q$. Hence, after sufficiently many time steps, the bracketed term will have suppressed the initial characteristic function almost everywhere except for a narrow region around $\phi = 0$. We can then perform a Taylor expansion in $|\phi| \ll 1$ to arrive at
\begin{align}
    \chi(\phi,k) &\approx \left[ 1 - p_\theta \frac{\phi^2}{2} + i v \phi + \mathcal{O} \{ \phi^3 \}  \right]^k \chi (\phi,0) = \left[ 1 - \frac{\phi^2}{2} \left(p_\theta-v^2 \right) + i v \phi  + \frac{1}{2} \left( i v \phi + \mathcal{O}\{\phi^{3/2}\} \right)^2  \right]^k \chi (\phi,0) \nonumber \\
    &\approx \exp \left[ - \frac{p_\theta-v^2}{2} k \phi^2 +i v k \phi \right] \chi (\phi,0)
\end{align}
Note that we have split the second order term in $\phi$ for a consistent expansion of the exponential function. As $\chi(\phi,k)$ will be narrowly peaked at small angles $\phi$, the corresponding charge distribution will be broad and thus admits a continuous description in $n$. Assuming an initial pure charge state at $n_0$, i.e.~$\chi(\phi,0) = e^{in_0\phi}$, and omitting the periodic boundary conditions, we are left with an approximately Gaussian charge distribution,
\begin{align} \label{eq:Pn_Gauss}
P(n,k) &= \int_{-\pi}^\pi \frac{\diff \phi}{2\pi} \chi(\phi,k) e^{-in\phi} \approx \int_{-\infty}^\infty \frac{\diff \phi}{2\pi} \chi(\phi,t) e^{-in\phi} \stackrel{k \gg 1}{\approx} \frac{1}{\sqrt{2\pi k ( p_\theta-v^2 )}} \exp \left[ -\frac{(n-n_0-v k)^2}{2k(p_\theta - v^2)} \right] %= \cN \left[ n_0 + vk, (p_\theta - v^2)k \right].
\end{align}
The normalization has to be corrected manually if the distribution is to be evaluated for discrete $n$. The mean and the variance of this Gaussian agree with the exact infinite-ladder results in \eqref{eq:classRWmoments}. A mixed initial charge state would result in a mixture of Gaussians, which would eventually converge towards a broader Gaussian. 

However, once the charge distribution hits the boundaries in the case of a finite battery, the Gaussian approximations no longer hold and the modified update rule for $P(0,k+1)$ and $P(N,k+1)$ will cause reflections and ultimately lead to a Gibbs-like steady state.

\section{Coherent battery charging}\label{app:infLadderQuantum}

We now consider a battery charged by qubits with energy coherences ($c>0$). Once again, the time evolution of the battery state has an exact analytic form so long as the charge boundaries are not yet occupied. To this end, we start again from the state transformation rule \eqref{eq:qubitChargeTrafo}, from which we see that the impact of the coherences is most prominent at half-swaps, $\theta = \pi/4$. The results presented in the main text were evaluated for for half-swaps and $c=1$, i.e.~with qubits prepared in the pure superposition state $\ket{\psi} = \sqrt{q}\ket{g} + \sqrt{1-q}\ket{e}$. 
In this case, we expect the coherences to cause interference effects as known from quantum random walks. In particular, a battery initialized in an intermediate charge state will evolve into a bimodal charge distribution as the charge steps accumulate, and the two branches of the distribution will simultaneously progress up and down the ladder until they hit the charge boundaries. 

The analytic solution to the battery evolution follows after taking the discrete Fourier transform of the density matrix with respect to $n,n'$, which formally amounts to switching into the periodic phase representation of the infinite-ladder Hilbert space, $\la \phi | \rho_B |\phi' \ra := \sum_{n,n'} \la n|\rho_B |n'\ra e^{in'\phi'-in\phi}/2\pi$. Introducing the short-hand notation  $\chi (\Phi, \varphi) := \la \Phi - \varphi/2 |\rho_B|\Phi + \varphi/2\ra$, \eqref{eq:qubitChargeTrafo} translates to
\begin{align}
    \la \phi | \rho_B' |\phi' \ra &= \left\{ 1-p_\theta + p_\theta \left[ (1-q)e^{i(\phi'-\phi)} + qe^{i(\phi-\phi')} \right] - ic \frac{\Omega}{2} \left[ e^{i\alpha} \left( e^{-i\phi} - e^{-i\phi'} \right) + e^{-i\alpha} \left( e^{i\phi} - e^{i\phi'} \right) \right] \right\} \la \phi | \rho_B |\phi' \ra \nonumber \\
    &= \left\{ 1 - p_\theta \left[ 1 - \cos(\phi'-\phi) \right] + i v \sin (\phi'-\phi)  - i c \Omega \left[ \cos (\phi-\alpha) - \cos (\phi' - \alpha) \right] \right\} \la \phi | \rho_B |\phi' \ra , \\
    \Rightarrow \chi' (\Phi,\varphi) &= \left[ 1 - p_\theta \left( 1 - \cos \varphi \right) + i v \sin \varphi  - 2 i c \Omega \sin (\Phi - \alpha) \sin \frac{\varphi}{2} \right] \chi (\Phi,\varphi). \nonumber
\end{align}
In this representation, a charge step merely amounts to a multiplicative factor. 
Notice that, since we employ identically prepared qubits with the same fixed phase angle $\alpha$, we are free to define the battery phase coordinates $\phi,\phi'$ relative to that reference, i.e.~set $\alpha=0$ without loss of generality.
The battery state after $k$ charge steps can now be expressed in terms of the initial state as 
\begin{align}
    \chi_k (\Phi,\varphi) &= \left[ 1 - p_\theta (1-\cos\varphi) + i v \sin\varphi - 2i c \Omega \sin \Phi \sin \frac{\varphi}{2} \right]^k \chi_0 (\Phi,\varphi). %\nonumber \\
    %& \added[id=n]{ = \left[ 1 +2i \sin \frac{\varphi}{2} \left( v \cos \frac{\varphi}{2} - c\Omega \sin \Phi \right) - 2p_\theta \sin^2 \frac{\varphi}{2} \right]^k \chi_0 (\Phi,\varphi). }
\end{align} 
One can check that the magnitude of the progression factor in square brackets is a number between zero and one, as it should be. It always assumes its maximum at $\varphi=0$ and decreases at first with growing $\varphi$. However, contrary to the incoherent case \eqref{eq:chiSol_class}, the magnitude does not necessarily stay substantially below one for all $(\Phi,\varphi)$-arguments. Indeed, at $\varphi = \pm \pi$ and $\Phi = \pm \pi/2$, the magnitude reaches up to $\sqrt{1-4p_\theta (1-p_\theta)[1-4c^2 q(1-q)]}$, which is unity for coherent qubit superpositions of equal weight ($c=1$, $q \approx 1/2$) and half-swaps ($p_\theta = 1/2$). Hence, substantial values of the characteristic function at large angles $|\varphi| \sim \pi$ can persist even after many charge steps, which explains why the charge distribution $P(n,k)$ may exhibit high-frequency interference fringes between neighbouring charge levels; 
%\added[id=n]{[CITE e.g.~quant-ph/0010117]}; 
the distribution can be expressed as the Fourier integral
\begin{align} \label{eq:Pn_chi_Fourier}
    P(n,k) &= \la n|\rho_B (k) |n \ra = \frac{1}{2\pi} \int_{-\pi}^{\pi}\diff \phi \diff \phi' e^{in(\phi - \phi')} \chi_k \left( \frac{\phi + \phi'}{2},\phi'-\phi \right) = \int_{-\pi}^\pi \frac{\diff \Phi}{2\pi} \int_{2|\Phi|-2\pi}^{2\pi - 2|\Phi|} \diff \varphi \, e^{-in\varphi} \chi_k (\Phi,\varphi). 
\end{align}
The high-frequency fringes seen in Fig.2 neither influence the distinct bimodal structure of the charge distribution after $k\gg 1$ nor do they affect the mean and variance appreciably. We shall therefore ignore those high-frequency components and focus on the coarse-grained, quasi-continuous distribution at long times, assuming that the relevant $\varphi$-values for the characteristic function are small.
Similar to the incoherent case, we can approximate consistently to 2nd order in $\varphi$ and are left with 
\begin{align}
    \chi_k (\Phi,\varphi) &\approx \left[ 1 + i\varphi \underbrace{\left( v -  c \Omega \sin \Phi \right)}_{=: \gamma (\Phi)} - p_\theta \frac{\varphi^2}{2} + \mathcal{O} (\varphi^3) \right]^k \chi_0 (\Phi,\varphi) \approx \exp \left[ - \frac{p_\theta - \gamma^2 (\Phi)}{2} k\varphi^2 + ik \gamma (\Phi) \varphi \right] \chi_0 (\Phi,\varphi). \label{eq:chi_Gaussianapprox}
\end{align}
So the characteristic function $\chi_k$ approaches a Gaussian shape in $\varphi$ whose width decreases like $1/\sqrt{k}$, while its complex phase oscillates at a frequency proportional to $k$; both depend on the other angle coordinate $\Phi$, as determined by 
\begin{equation}\label{eq:gammaPhi}
    \gamma(\Phi) = v - c\Omega  \sin\Phi = (1-2q) \sin^2 \theta + c\sqrt{q(1-q)} \sin 2\theta \sin \Phi.
\end{equation}
The charge distribution \eqref{eq:Pn_chi_Fourier} can be seen as the $[n-k\gamma(\Phi)]$-th Fourier component with respect to $\varphi$, truncated and averaged over $\Phi$. For large $k$, the $\Phi$-average will be dominated by the vicinity of the points $\Phi_\pm = \pm \pi/2$ at which the strongly oscillating phase is stationary. Moreover, the width of the Gaussian will be much smaller than $\pi$, which allows us to extend the $\varphi$-integral in \eqref{eq:Pn_chi_Fourier} to infinity. Inserting \eqref{eq:chi_Gaussianapprox} and assuming that the battery is initially prepared in a charge eigenstate $|n_0\ra$ with $\chi_0 (\Phi,\varphi) = e^{in_0 \varphi}/2\pi$, we get
\begin{align}
    P(n,k) &\approx \frac{1}{4\pi^2} \int_{-\pi}^\pi \diff \Phi \int_{-\infty}^{\infty} \diff \varphi \, \exp \left\{ - \frac{p_\theta - \gamma^2 (\Phi)}{2} k\varphi^2 - i[n- n_0 - k \gamma (\Phi)] \varphi \right\} \nonumber \\
    &= \frac{1}{2\pi} \int_{-\pi}^\pi \diff \Phi \frac{1}{\sqrt{2\pi k[p_\theta - \gamma^2 (\Phi)]}} \exp \left\{-\frac{[n - n_0 - k\gamma(\Phi)]^2}{2k [p_\theta - \gamma^2 (\Phi)] }\right\} .
    \label{eq:Pn_Gauss_coh}
\end{align}
The result is a smooth average over Gaussians with large widths and displacements. A mixed initial charge state would entail an additional weighted sum over different $n_0$, accordingly. 

The charge moments are consistently evaluated in the continuum approximation, as we are assuming that the charge distribution \eqref{eq:Pn_Gauss_coh} be broad, $\overline{n^j} (k) = \sum_n n^j P(n,k) \approx \int\diff n\, n^j P(n,k)$.
Specifically, we get
\begin{align}
    \overline{n} (k) & \approx n_0 + k \frac{1}{2\pi} \int_{-\pi}^\pi \diff \Phi \, \gamma(\Phi) = n_0 + vk , \\
    \overline{n^2} (k) & \approx \frac{1}{2\pi} \int_{-\pi}^\pi \diff \Phi \, \left[ k p_\theta - k\gamma^2 (\Phi) + (k\gamma(\Phi) + n_0)^2 \right] = n_0^2 + k (p_\theta + 2v) + k(k-1) \left(v^2 + \frac{c^2 \Omega^2}{2} \right), \nonumber \\
    \Delta n^2 (k) &\approx (p_\theta - v^2)k + \frac{c^2\Omega^2}{2} k(k-1), \nonumber 
\end{align}
see also Eq.9 in the main text. Note that the charge variance only holds for a \emph{pure} initial state, and one must add the initial spread $\Delta n^2(0)$ in the case of a mixture. The average charge moves at the same speed as in the incoherent charging case; there is no difference within the validity bounds of our approximations. The variance, however, gets an additional contribution that grows quadratically with $k$ and thus eventually becomes the the dominant cause of charge fluctuations. Indeed, our numerical simulations show that the asymptotic steady state of a coherently charged battery typically yields a flat energy distribution that extends over the whole ladder.

Finally, in order to see that \eqref{eq:Pn_Gauss_coh} describes a bimodal distribution, recall that the displacement parameter $\gamma(\Phi)$ in \eqref{eq:gammaPhi} oscillates between its two extreme values, $\gamma (\Phi_{\pm}) =  v \mp c\Omega$. Since these are stationary points, the $\Phi$-integral will allocate most weight to them and result in distinct maxima around the two respective mean charges $n_{\pm} \approx n_0 + vk \pm c\Omega k$. 
We can make this explicit by performing a stationary phase approximation with respect to $\Phi$ in the first line of \eqref{eq:Pn_Gauss_coh}. Consider two well-behaved real-valued functions $f(x), g(x)$ for $x\in [a,b]$ and let $x_n$ be non-degenerate stationary points well within that interval, at which $f'(x_n) = 0$ and $f''(x_n) \neq 0$. Then 
\begin{equation}
    \int_a^b \diff x\, g(x) e^{ik f(x)} \approx \sum_{x_n} \sqrt{\frac{2\pi i }{k f''(x_n)}} g(x_n) e^{ikf(x_n)} \qquad \text{for }\,\, k\to\infty.
\end{equation}
Applied to \eqref{eq:Pn_Gauss_coh}, the approximation is only strictly justified for $|\varphi| \gg 1/k$ and should therefore be taken as a qualitative estimate that may well lead to deviations in the flat parts of the charge distribution while reproducing its more sharply peaked features. We arrive at
\begin{align}
    P(n,k) &\approx \frac{1}{4\pi^2} \int_{-\infty}^{\infty} \diff \varphi \left\{ \sqrt{\frac{2\pi i}{c\Omega k \varphi}} \exp \left[ - \frac{p_\theta - \gamma^2(\Phi_+)}{2} k \varphi^2 + ik\gamma(\Phi_+) \varphi \right] \right. \nonumber \\ &+ \left. \sqrt{-\frac{2\pi i}{c\Omega k \varphi}} \exp \left[ - \frac{p_\theta - \gamma^2(\Phi_-)}{2} k \varphi^2 + ik\gamma(\Phi_-) \varphi \right] \right\} e^{-i(n-n_0)\varphi} \nonumber \\
    &= \frac{Q_+ \left[ n-n_0-k\gamma(\Phi_+), p_\theta - \gamma^2(\Phi_+) \right] + Q_- \left[ n-n_0-k\gamma(\Phi_-), p_\theta - \gamma^2(\Phi_-) \right]}{\sqrt{4c \Omega k}}, \label{eq:Pn_statPhase_coh}
\end{align}
introducing a distinctively peaked, unnormalized combination of Gaussian distribution and modified Bessel functions,
\begin{align}
    Q_\pm [\mu,\sigma^2] &= \frac{1}{\sqrt{4\pi \sigma^2}}e^{-\mu^2/4\sigma^2} \left[ \sqrt{|\mu|} \,  I_{-1/4} \left( \frac{\mu^2}{4\sigma^2} \right) \pm \frac{\mu}{\sqrt{|\mu|}} I_{1/4} \left( \frac{\mu^2}{4\sigma^2} \right) \right].
\end{align}
Although the stationary phase approximation \eqref{eq:Pn_statPhase_coh} is noticably less accurate than the Gaussian approximation \eqref{eq:Pn_Gauss_coh}, it makes the double-peak structure of the charge distribution around $n_\pm$ explicit.

\section{Charging efficiency using free energy}
\begin{figure}
\centering
\includegraphics[width=0.8\columnwidth]{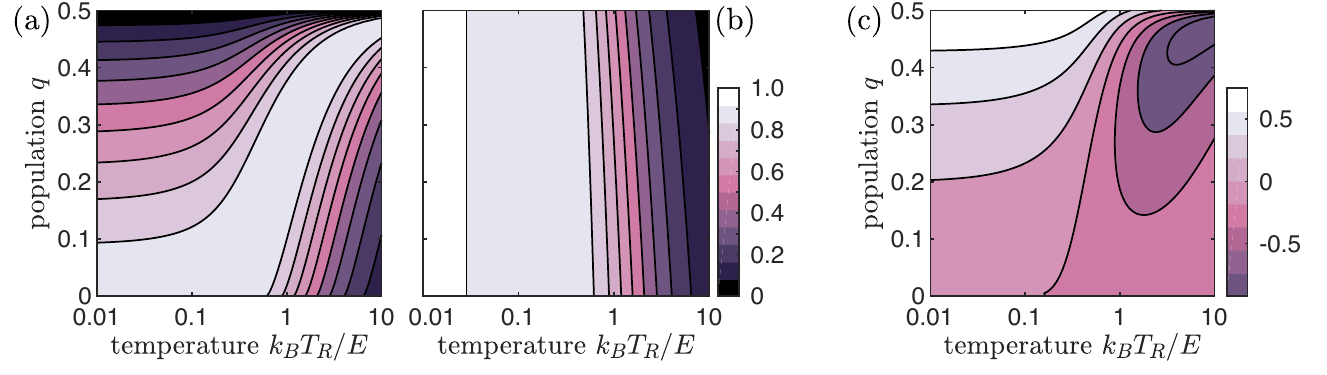}
\caption{\label{fig:compareFreeEnergy} Ratio of ergotropy $\cE_Q$ to free energy difference $\Delta\cF_Q(T_R)$ for (a) incoherent ($c=0$) and (b) coherent ($c=1$) qubits at different $q$ and reference temperature $T_R$ used to prepare the qubits. Panel (c) shows the difference in ratio $\cE_Q^{c=1}/\Delta\cF_Q^{c=1}(T_R)  - \cE_Q^{c=0}/\Delta\cF_Q^{c=0}(T_R)  $.}
\end{figure}

In the main text, we defined the average efficiency after a given number of charging steps by the total ergotropy stored in the battery over the total ergotropy provided by the qubits, $\eta(k) = \cE_B(k)/k\cE_Q$ with 
\begin{equation}
    \cE_Q = \frac{E}{2} \left[1-2q + \sqrt{(1-2q)^2 + 4c^2q(1-q)} \right].
\end{equation}
The definition implies that ergotropy is the relevant input resource, i.e.~qubits in passive zero-ergotropy states of a given von Neumann entropy $\cS(\rho_Q) = -\tr\{ \rho_Q \ln \rho_Q \}$ are freely available and the charging cost is the energy required to prepare the population-inverted and/or coherent qubit state \eqref{eq:qubitState} by means of a unitary (isentropic) operation. 
Alternatively, one could assume that the same qubit state was prepared in an isothermal process at a given reference temperature $T_R$ starting from equilibrium, in which case the required energy cost would be at least the free energy difference to the Gibbs state, 
\begin{equation}
    \Delta \cF_Q (T_R) = E(1-q) - k_B T_R \left[ \cS(\rho_Q) - \ln \left(1+e^{-E/k_B T_R} \right) \right] = \cF_Q(T_R) + k_B T_R  \ln \left(1+e^{-E/k_B T_R} \right) \geq \cF_Q (T_R).
\end{equation}
The corresponding efficiency would read as $\eta(k,T_R) = \cE_B(k)/k \Delta \cF_Q (T_R)$.

In the case of coherent charging ($c=1)$, the qubit is in a pure state with $\cS(\rho_Q) = 0$, for which the ergotropy and the free energy content match, $\cE_Q = \cF_Q (T_R) = E(1-q)$. For incoherent charging with population-inverted qubits ($c=0$ and $q<1/2$), on the other hand, we have $\cE_Q = E(1-2q)$ and $\cF_Q (T_R) = E(1-q) + k_B T_R [q\ln q + (1-q)\ln(1-q)]$. Hence, the free energy resources in both cases are
\begin{align}
    \Delta\cF_Q^{(c=1)}(T_R) &= E(1-q) + k_B T_R \ln \left(1+e^{-E/k_B T_R} \right) \geq E(1-q) = \cE_Q^{(c=1)}, \\
    \Delta\cF_Q^{(c=0)}(T_R) &= E(1-q) + k_B T_R \left[ q\ln q + (1-q)\ln (1-q) + \ln \left(1+e^{-E/k_B T_R} \right) \right] \geq E(1-2q) = \cE_Q^{(c=0)}.
\end{align}
The inequality in the first line is trivial, while the second inequality can be checked numerically. Both inequalities imply $\eta(k) > \eta(k,T_R)$, and they align with the fact  that the free energy gives the upper bound to the maximum energy extractable from a state at a given temperature (see e.g. Refs.~\cite{Skrzypczyk2014,Bera2019}).

Figure \ref{fig:compareFreeEnergy} shows the ratio of $\cE_Q$ to $\Delta \cF_Q (T_R)$, or equivalently of $\eta(k,T_R)$ to $\eta(k)$, at different reference temperatures $T_R$ and populations $q$ for incoherent ($c=0$) and coherent ($c=1$) charging qubits. 
In the case of incoherent charging, the ratio drops with increasing $q$ and $T_R$ (while remaining high when both simultaneously increase), whereas for the coherent case, the ratio remains close to one up until the high-temperature regime $ k_B T_R > E$.

In the main text, we use the ergotropy rather than the free energy difference as a figure of merit for quantifying efficiency, because it does not require a reference temperature $T_R$ and only depends on the state. While this may over-predict the net efficiency of the charging process with thermodynamically prepared qubits, the advantage of coherent over incoherent charging will persist at low reference temperatures $k_B T<0.15E$ since $\cE_Q^{c=1}/\Delta\cF^{c=1}\gtrsim 0.8$ for all $q$-values (see Fig.~\ref{fig:compareFreeEnergy}(c)) while $\cE_Q^{c=0}/\Delta\cF^{c=0}$ ranges from $0$ to $1$ for different $q$-values.

The comparison of charging efficiency in terms of ergotropy and in terms of free energy difference to a reference bath highlights a subtlety on the precise thermodynamic resource that causes the quantum enhancement: is the greater charging power/efficiency at fixed $q$ due to the increased purity (i.e.~lower von Neumann entropy) or the increased coherence of the charge qubits? Our analysis suggests the latter, because it is the presence of coherence that causes the buildup of interference effects in the battery, which in turn lead to the described bimodal energy distribution and the coherent charging speed-up. Low-entropy qubit states without energy coherence, on the other hand, would never generate such interference effects. Specifically, we discuss in the main text that quantum coherent states can beat arbitrary classical strategies including those using pure excited-state qubits in terms of charging power.
%\textbf{The use of free energy in quantifying efficiency suggests the von Neumann entropy as a possible thermodynamic resource. However, we attribute the enhancement in battery charging to coherence rather than purity/entropy of the state. This is because the bimodal distribution for coherent charging is a direct consequence of the amount of coherence in the qubits, where the drift velocities speed up and slow down by $c\Omega$. Indeed, the presence of coherence is crucial in order to buildup interference effects in the battery (low-entropy but non-coherent states would never generate such interference effects), which in turn leads to a speed-up of the fast-moving peak. Hence, we conclude that the speed-ups are directly caused by the presence of coherence. Furthermore, we discuss in the main text that quantum coherent states can beat classical strategies including the use of pure excited-state qubits.}
\end{widetext}

\end{document}